\documentclass[3p,times,procedia]{elsarticle}
\usepackage{nupha_ecrc}

\volume{00}

\firstpage{1}

\journalname{Nuclear Physics A}

\runauth{R. Arnaldi}

\jid{nupha}

\jnltitlelogo{Nuclear Physics A}

\usepackage{amssymb}

\usepackage[figuresright]{rotating}

\newcommand{\jpsi}{\rm J/$\psi$}
\newcommand{\psip}{$\psi(\rm 2S)$}
\newcommand{\upsi}{$\Upsilon$}
\newcommand{\sqrts}{$\sqrt{\rm s}$ }
\newcommand{\sqrtsNN}{$\sqrt{\rm s_{NN}}$ }
\newcommand{\raa}{$R_{\rm AA}$}
\newcommand{\rpa}{$R_{\rm pA}$ }
\newcommand{\smallraa}{$r_{\rm AA}$}
\newcommand{\pt}{$p_{\rm T}$}
\newcommand{\cc}{$c\bar{c}$ }
\newcommand{\npart}{$N_{\rm part}$}

\begin{document}

\begin{frontmatter}



\dochead{}

\title{Experimental overview on quarkonium production}


\author{R. Arnaldi}

\address{INFN Torino, Via P. Giuria 1, I-10125 Torino (Italy)}

\begin{abstract}
Quarkonium production in heavy-ion collisions is a well-known signature of the formation of a plasma of quarks and gluons (QGP). After thirty years from the first measurements at SPS energies, a large wealth of results is now accessible from high-energy experiments at RHIC and LHC, and these new data are contributing to sharpen the picture of the quarkonium behaviour in \mbox{A-A} collisions.  
In this paper, an overview of the main results on both charmonium and bottomonium production in \mbox{p-A} and \mbox{A-A} collisions is presented, focussing on the most recent achievements from the RHIC and LHC experiments.
\end{abstract}

\begin{keyword}
Quarkonium \sep Heavy-ion collisions \sep Quark-Gluon Plasma


\end{keyword}

\end{frontmatter}


\section{Introduction}
\label{sec:intro}
Quarkonium is a bound state of a Q and ${\bar{\rm Q}}$ pair, where ${\rm Q}$ can be either a charm quark (therefore forming a charmonium state) or a bottom quark (bottomonium state). 
Since thirty years, quarkonia are considered important probes of the formation,
in heavy-ion collisions, of a strongly interacting medium, the so-called quark gluon plasma (QGP).
In a hot and deconfined medium, in fact, quarkonium production is expected to be significantly suppressed
with respect to the proton-proton yield, scaled by the number of binary nucleon-nucleon
collisions. The origin of such a suppression, taking place in the QGP, is the color screening of
the force which binds the $c\bar{c}$ ($b\bar{b}$) state~\cite{Matsui:1986dk}. In this scenario, quarkonium suppression should occur
sequentially, according to the binding energy of each meson. In the charmonium sector strongly bound states, as the $\rm J/\psi$, should
melt at higher temperatures with respect to the more loosely bound $\chi_{c}$ or \psip\ resonances, while in the bottomonium sector, the \upsi(1S) is expected to melt at a higher temperature with respect to the $\chi_{b}$ or \upsi(2S) and \upsi(3S). 
As a consequence, the in-medium dissociation probability of these states should provide an estimate of the initial temperature reached in the collisions~\cite{Digal:2001ue}. However, the connection between the dissociation temperature and the disappearance of a given quarkonium state is not yet straightforward. From the theory point of view, large uncertainites are still associated to the evaluation of such a temperature, depending on the adopted approach as lattice quantum chromodynamics (QCD), QCD sum rules, AdS/QCD, resummed perturbation theory, effective field theories or potential models~\cite{Adare:2014hje}. From the experimental point of view, the prediction of a sequential suppression pattern is complicated by several factors as the feed-down contributions from higher-mass resonances into the observed quarkonium yield and the B-hadrons decay into charmonium. Furthermore, other
hot and cold matter effects can play a role, competing with the suppression mechanism.
Increasing the center of mass energy of the collisions (\sqrts), an increase of the production of ${\rm Q}$ and $\bar{\rm Q}$ quarks is expected. Therefore, in high-\sqrtsNN\ interactions the abundance of ${\rm Q}$ and $\bar{\rm Q}$ quarks might lead to a new quarkonium production source, originating from the (re)combination of these quarks during the collision history~\cite{Thews:2000rj} or at the hadronization~\cite{BraunMunzinger:2000px,Andronic:2011yq}. This additional production mechanism is related to the formation of a hot medium and is particularly important for charmonium states, given the higher number of \cc\ with respect to $b\bar{b}$\ pairs. (Re)combination is expected to enhance the \jpsi\ yields, therefore smoothening or even counterbalancing the effect of the suppression mechanism.
Quarkonium is also expected to be affected by several effects related to cold matter
(the so-called cold nuclear matter effects, CNM). For example, the production cross section of the
${\rm Q}\bar{\rm Q}$ pair is influenced by the kinematic distributions of partons in the nuclei, which are different from
those in free protons and neutrons (this effect is known as nuclear shadowing~\cite{Eskola:2009uj,deFlorian:2011fp,Hirai:2007sx}). 
In a similar way, approaches based on the Color-Glass Condensate effective theory~\cite{Kharzeev:2005zr,Fujii:2006ab} assume that gluon saturation is expected to set in at high energies. This effect influences the quarkonium production occurring through fusion of gluons carrying small values of the Bjorxen-$x$ in the nuclear target.
Furthermore, parton energy loss may decrease the pair momentum~\cite{Vogt:1999dw,Arleo:2012hn}, causing a reduction of
the quarkonium production at large longitudinal momentum. Finally, while the ${\rm Q}\bar{\rm Q}$ pair evolves towards
the final quarkonium state, it may also interact with the medium and eventually break-up. This
effect is expected to play a dominant role only for low-\sqrtsNN collisions, where the crossing time of the
(pre)-resonant state in the nuclear environement is rather large.
Cold nuclear matter effects are investigated in proton-nucleus collisions, where no hot medium is
expected to be formed. Since these effects are also present in nucleus-nucleus interactions, a precise
knowledge of their role is crucial to correctly quantify the influence of the hot QCD medium on quarkonium. 

The in-medium modification of quarkonium production, induced by either hot or cold matter effects, is evaluated  through the nuclear modification factor
\raa, defined as the ratio of the quarkonium yield in \mbox{A-A} collisions ($Y_{\rm AA}^{q\overline{q}}$)
and the expected value 
obtained by scaling the \mbox{pp} yield ($Y_{\rm pp}^{q\overline{q}}$) by the average number of nucleon-nucleon collisions, $\langle N_{\rm{coll}}\rangle$, evaluated through a Glauber model calculation:
\begin{center}
$R_{\rm AA} = Y_{\rm AA}^{q\overline{q}}/(\langle N_{\rm coll} \rangle \times Y_{\rm pp}^{q\overline{q}})$
\end{center}
 
\raa is expected to be equal to unity if the quarkonium yield in \mbox{A-A} scales with $N_{\rm coll}$, as it is the case for electroweak probes (direct $\gamma$, W, Z) which do not interact strongly.
On the contrary, \raa\ different from unity might imply that the quarkonium production is affected by the medium.

The ``history'' of quarkonium studies in heavy-ion collisions dates back to 30 years ago, when the first measurements were performed at the CERN SPS at \sqrtsNN=17 GeV. This first measurements, limited only to the charmonium family~\cite{Abreu:2000ni,Arnaldi:2007zz}, provide an indication for a clear suppression of the \jpsi\ meson (the so-called ``anomalous \jpsi\ suppression''), well beyond the expectations of cold nuclear matter effects. 
Starting from year 2000, charmonium and first bottomonium results were achieved also at RHIC at the top energy of \sqrtsNN=200 GeV. In the following years, the RHIC quarkonium program focussed on the investigation of the quarkonium production by varying the colliding beam species and the energy of the collisions. Results obtained at \sqrtsNN=200 GeV indicate a \jpsi\ suppression similar to the one observed at SPS, although the factor $\sim$10 increase in center of mass energy~\cite{Adare:2006ns,Adare:2011yf}.
Furthermore, unexpectedly, a stronger \jpsi\ suppression has been
measured at RHIC at forward with respect to mid-rapidity ($y$), in spite of the higher energy density
which is reached close to $y\sim0$. These observations suggest the existence of the previously mentioned
(re)combination process, which might set in already at RHIC energies and which may counteract the
quarkonium suppression in the QGP.
At LHC, ALICE, ATLAS and CMS eventually carried out studies on quarkonium in \mbox{Pb-Pb} collisions at the center of mass energy of \sqrtsNN=2.76 TeV (LHC Run-I). 
The high-energy density of the medium and the large number of $c\bar{c}$ pairs produced in central \mbox{Pb-Pb} collisions make measurements at LHC energy an excellent testing ground for suppression and (re)combination scenarios. Furthermore, the study of the bottomonium, now largely accessible due to the \sqrts-increase of the $b$ production cross-section, allows a deeper understanding on quarkonium behaviour in \mbox{A-A}.

These experiments are characterized by different kinematic coverages, allowing the investigation of quarkonium production over $\sim$4 rapidity units, from 0 to high transverse momentum (\pt).
ATLAS and CMS are designed to measure quarkonium production reconstructing
the various states in their dimuon decay channel. They both cover the mid-rapidity region: depending
on the quarkonium state under study and on the \pt\ range investigated, the CMS rapidity coverage
can reach up to $|y|<$2.4, and a similar $y$ range is also covered by ATLAS. ALICE measures
quarkonium in two rapidity regions: at mid-rapidity ($|y|<$0.9) in the dielectron decay channel and 
at forward rapidity (2.5$<y<$4) in the dimuon decay channel, in both cases down to zero transverse
momentum. LHCb has taken part only to the pp and p-A LHC programs \footnote{In Fall 2015, LHCb has, for the first time,  taken part to the LHC \mbox{Pb-Pb} data taking (LHC Run-II).} and results on quarkonium
production, reconstructed through the dimuon decay channel, are provided at forward rapidity, i.e.
2$<y<$4.5.

It has to be noted that at the three facilities where quarkonium in \mbox{A-A} collisions has been studied, a complementary experimental \mbox{pp}, \mbox{p-A} or \mbox{d-A} program has always been foreseen, meant to address quarkonium production in smaller systems, but also to provide a reference for \mbox{A-A} studies.

In the following sections, the latest experimental quarkonium results presented at the Quark Matter 2015 Conference will be discussed, focussing on the \mbox{A-A} and \mbox{p-A} measurements.

%

\section{Quarkonium in \mbox{A-A} collisions}
\label{sec:AA}
\begin{figure}[h]
\centering
\includegraphics[width=7cm,clip]{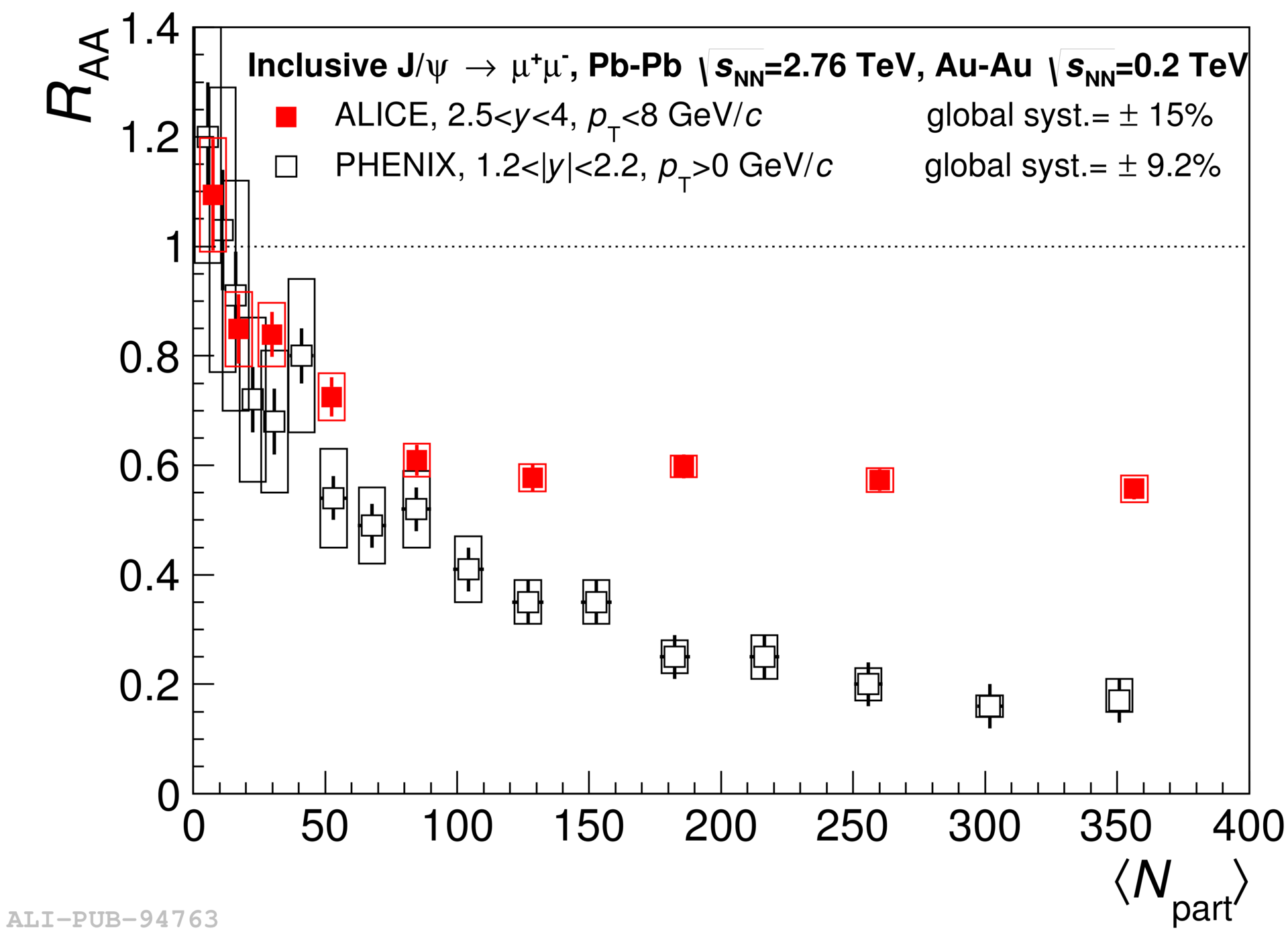}
\includegraphics[width=5.5cm,clip]{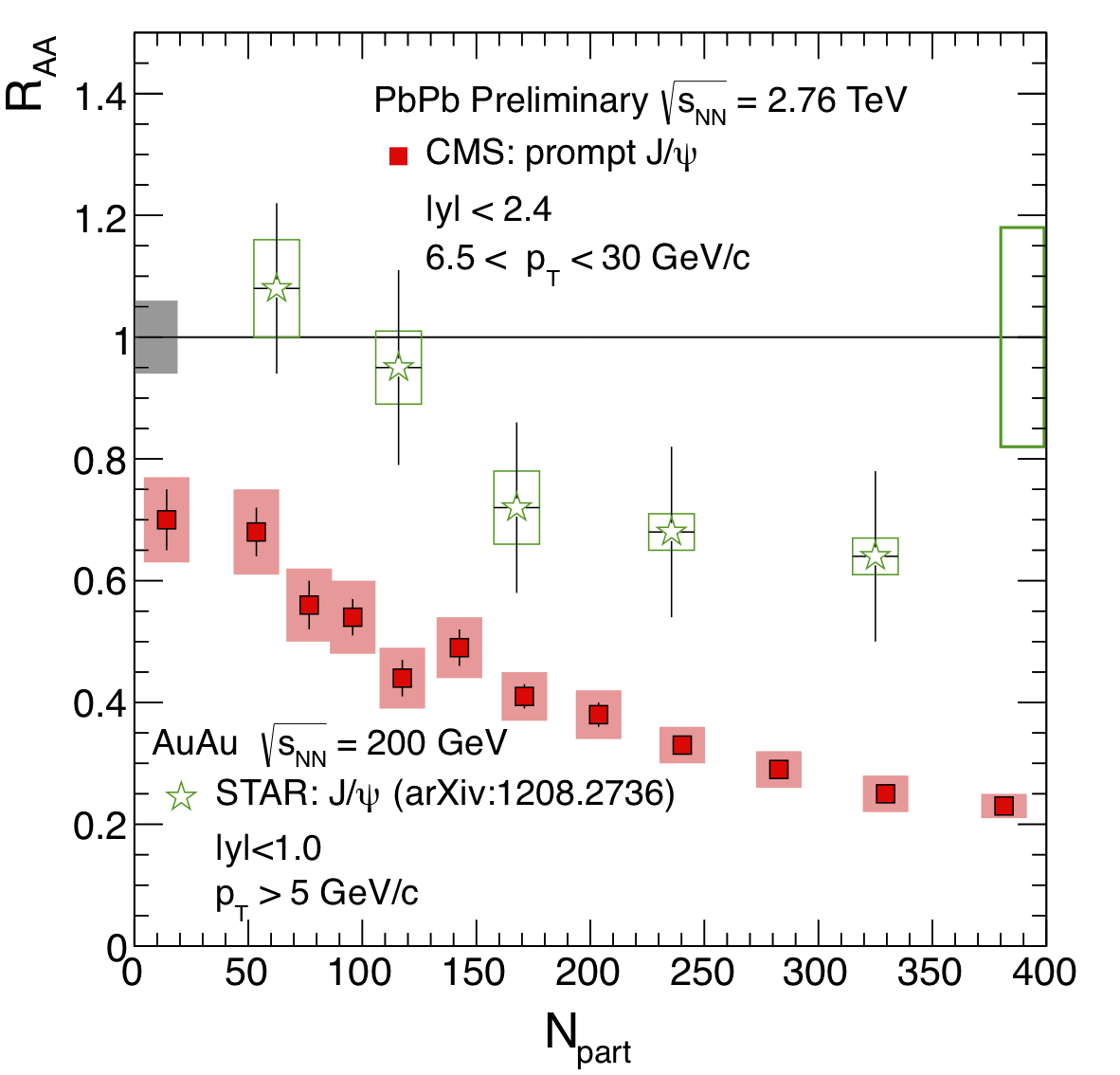}
\caption{Left: ALICE~\cite{Abelev:2013ila} and PHENIX~\cite{Adare:2011yf} inclusive low \pt\ \jpsi\ nuclear modification factor versus the number of participant nucleons at forward rapidity. Right: CMS~\cite{CMS:2012wba} and STAR~\cite{Adamczyk:2012ey} high-\pt\ \jpsi\ \raa\ versus \npart.}
\label{fig:RAAJPsi}       
\end{figure}

ALICE has studied the centrality dependence of the inclusive \jpsi\ (prompt \jpsi\  plus those coming from
B-hadron decays) production at low \pt\, in \mbox{Pb-Pb} collisions~\cite{Abelev:2013ila,Adam:2015isa}. The corresponding nuclear modification factor, as a function of the number of participant nucleons ($N_{part}$), is shown in Fig.~\ref{fig:RAAJPsi} (left) where it is compared to the \raa measured by the PHENIX experiment~\cite{Adare:2011yf} in a similar kinematic range. 
There is a clear evidence for a smaller \jpsi\ suppression at LHC energies, with respect to RHIC. Furthermore, while PHENIX result shows an increasing \jpsi\ suppression towards the most central events, the ALICE \jpsi\ \raa    saturates for \npart$>$100. Partonic transport models, which include, as \jpsi\ production process,  
the (re)combination of \cc pairs along the history of the collisions, indeed predict such a behaviour~\cite{Zhao:2011cv,Liu:2009nb}. The smaller suppression at the LHC is due to the larger \cc pair multiplicity which allows a new production source to set in, resulting in a compensation of the suppression from color screening. A similar behaviour is expected by the statistical model~\cite{Andronic:2011yq}, where the \jpsi\  yield is completely determined by the chemical freeze-out conditions and by the abundance of \cc pairs. A more precise comparison between the experimental result and the theory predictions would clearly benefit from an experimental measurement of the charm production cross-section, which represents one of the main sources of theory uncertainties. 

The (re)combination process is expected to be dominant in central collisions and, for kinematical reasons, it should contribute mainly at low \pt, becoming negligible as the \jpsi\ \pt\ increases.
This observation is confirmed by the mid-rapidity \raa\ measured by CMS~\cite{CMS:2012wba} and STAR~\cite{Adamczyk:2012ey} for high-\pt\ \jpsi, i.e. for \jpsi\ in a kinematic region where the (re)combination contribution can be neglected. As shown in Fig.~\ref{fig:RAAJPsi}(right), the \jpsi\ suppression, in this case, is stronger at LHC energies, as expected from a QGP dissociation scenario. The mid-rapidity STAR \raa\ measurement is also confirmed by the new results obtained at forward $y$~\cite{Ma_QM15}.

Another hint for the \pt\ dependence of the (re)combination process can be obtained studying the \smallraa, defined as the ratio of the measured $<p_{\rm T}^{2}>$ in \mbox{A-A} and in \mbox{pp} collisions. While at SPS or RHIC energies, the \smallraa\ shows an increasing pattern towards high centralities, ALICE~\cite{Pereira_QM15,Adam:2015isa} observes a decreasing \smallraa\ trend related to the enhanced \jpsi\ production at low \pt\ due to (re)combination, as confirmed by transport models~\cite{Zhao:2011cv,Zhou:2014kka} already reproducing the \raa\ evolution. 

The energy increase by a factor $\sim$2 achieved in LHC Run-II in Fall 2015 (\mbox{Pb-Pb} collisions at \sqrtsNN=5.02 TeV) together with the higher luminosity, are expected to shed more light on the (re)combination and suppression scenarioes.

A new result presented at this Conference by ALICE concerns the unexpected strong enhancement observed in the  \raa~\cite{Adam:2015gba,Martinez_QM15} for very low \pt\ \jpsi\ (0$<$\pt$<$0.3 GeV/c). This excess reaches a significance larger than 5 in the most peripheral collisions (70-90\%) and it might be due to a coherent \jpsi\ photoproduction, observed, for the first time, in \mbox{A-A} collisions. Also in this case, the higher luminosity reached in LHC Run-II will clearly help to improve the precision of such a measurement.

\begin{figure}[h]
\centering
\includegraphics[width=7.3cm,clip]{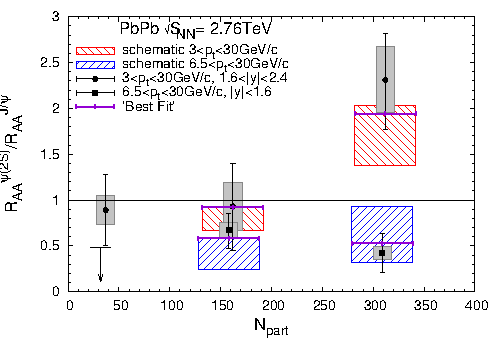}
\includegraphics[width=7cm,clip]{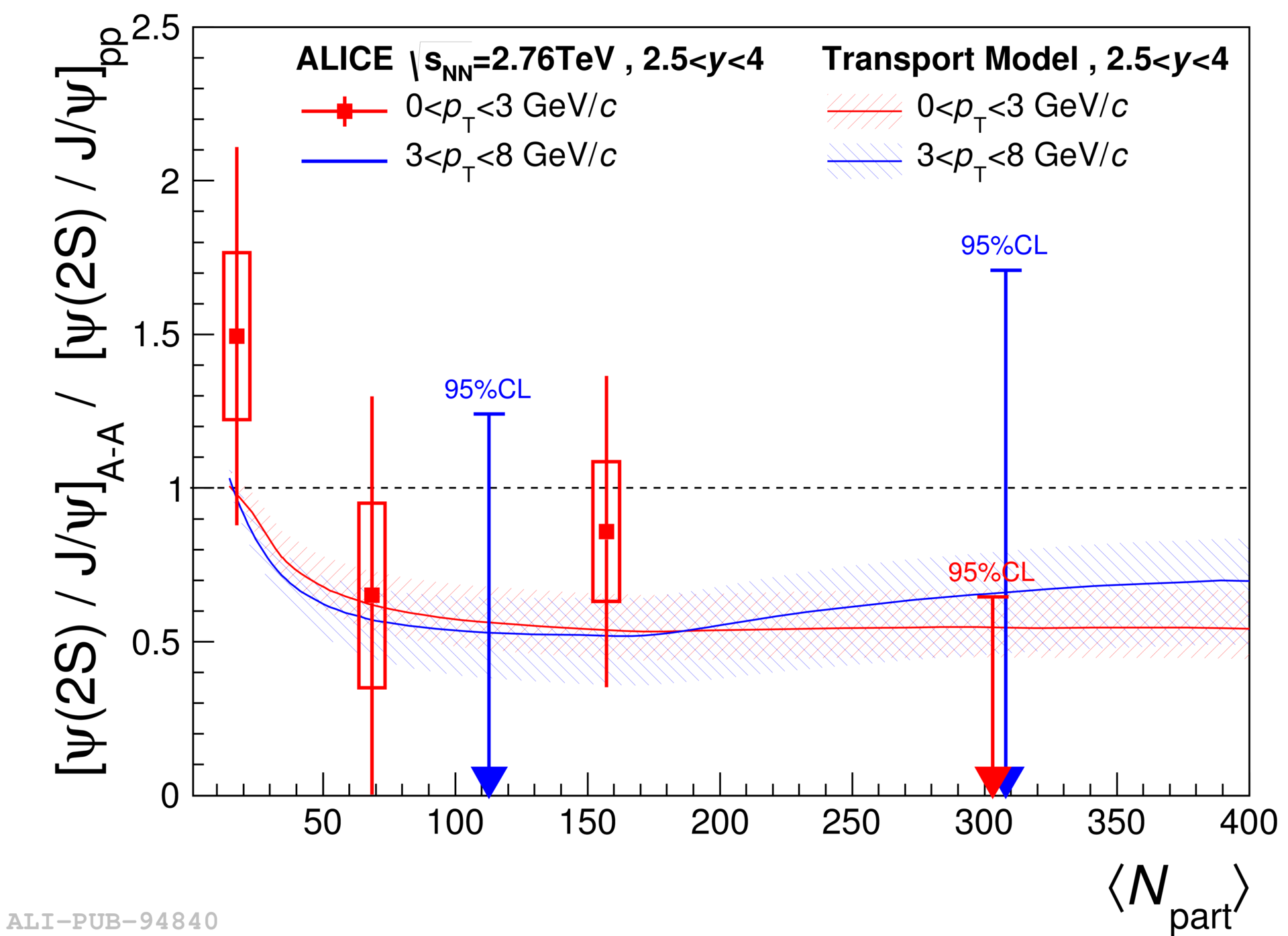}
\caption{\psip\ to \jpsi\ double ratio measured by CMS~\cite{Khachatryan:2014bva,Kim_QM15} (left) and ALICE~\cite{Adam:2015isa,Pereira_QM15} (right), as a function of \npart, compared to transport models~\cite{Du:2015wha,Chen:2013wmr}.}
\label{fig:psipAA}       
\end{figure}

Further insight on charmonium production in \mbox{Pb-Pb} collisions can be achieved comparing the 
production yields of a higher mass resonance, as the \psip, to the \jpsi\ ones. 
Results, presented as a double ratio of the  \psip\ to \jpsi\ 
yields in \mbox{Pb-Pb} and in \mbox{pp} collisions as a function of centrality,
are shown in Fig.\ref{fig:psipAA} for CMS~\cite{Khachatryan:2014bva,Kim_QM15} (left) and for  ALICE~\cite{Adam:2015isa,Pereira_QM15} (right). 
CMS observes values higher than one in the region 1.6$<|y|<$2.4, 3$<$\pt$<$30 GeV/$c$, implying a \psip\ \raa\ larger than the \jpsi\ one. Later \psip\ (re)generation, when radial flow is stronger, might explain the observed rise~\cite{Du:2015wha}. Results obtained in the range $|y|<$1.6 and 6.5$<$\pt$<$30 GeV/$c$ show a decreasing pattern towards most central collisions, i.e. the \psip\ \raa\ is, in this case, smaller than the corresponding \jpsi\ value, as expected in a sequential suppression scenario.
ALICE explores a contiguous range in rapidity (2.5$<y<$4) and also extend the 
\pt\ reach of this measurement down to zero, showing a trend in agreement with transport models~\cite{Chen:2013wmr} and with the statistical hadronization approach~\cite{Andronic:2009sv}.
Unfortunately, the large statistics and systematic uncertainties associated to the \psip\ measurement preclude the drawing of strong conclusions on the \psip\ behavior and on its kinematic dependence. Also this measurement will surely benefit from the high quarkonium statistics collected in LHC Run-II.

\begin{figure}[h]
\centering
\includegraphics[width=6cm,clip]{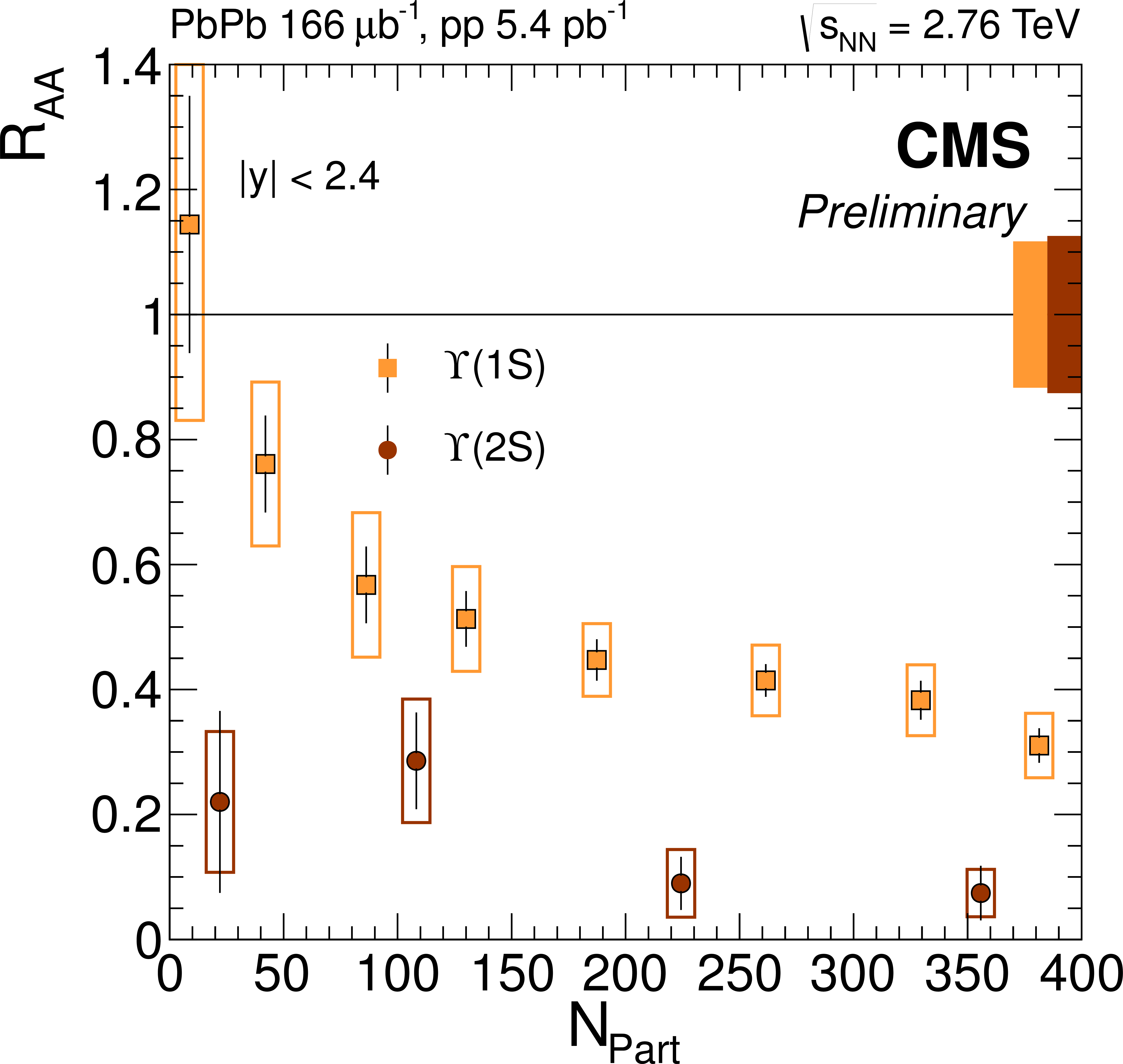}
\includegraphics[width=5.75cm,clip]{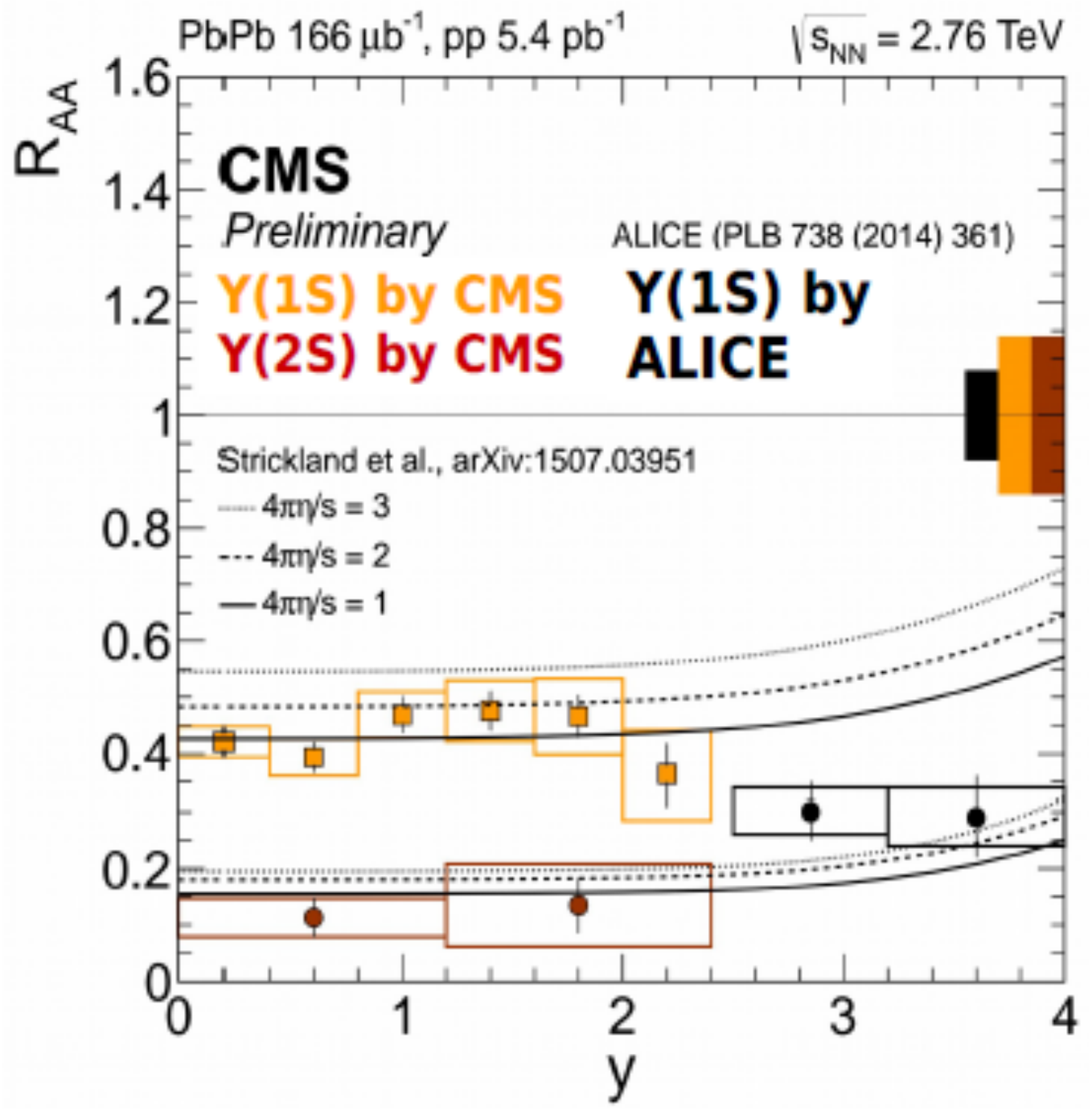}
\caption{Left: \raa\ of \upsi\ states as a function of centrality, as measured by  CMS~\cite{Jo_QM15,CMS:PASHIN15}. Right: \upsi\ \raa\ as a function of rapidity, as measured by CMS~\cite{Jo_QM15,CMS:PASHIN15} and ALICE~\cite{Das_QM15,Abelev:2014nua}.}
\label{fig:upsiAA}       
\end{figure}

Finally, the highest LHC energies allow the study of the resonances of the bottomonium family. New CMS results~\cite{Jo_QM15,CMS:PASHIN15}, based on an improved muon reconstruction algorithm and on a high-statistics \mbox{pp} reference, confirm the sequential suppression observed for \upsi(1S) 
(\raa= 0.43$\pm$0.03$\pm$ 0.07), \upsi(2S) (\raa= 0.13$\pm$0.03$\pm$0.02) and \upsi(3S) (\raa= 0.14 at 95\% CL.), as shown in Fig.~\ref{fig:upsiAA} (left).
Recent measurements of feed-down from excited states (for a review see~\cite{HermineQWG14}) seem not enough to explain the observed \upsi(1S) suppression, pointing to a possible suppression also for this strongly-bound state.
While the \upsi\ \raa\ show a clear centrality dependence, a rather flat trend is observed as a function of \pt\ or $y$, as shown in Fig.~\ref{fig:upsiAA} (right). The \raa\ results versus \pt\ and in a large rapidity coverage, extended up to 4 with the ALICE measurement~\cite{Das_QM15,Abelev:2014nua}, provide strong constraints for the theoretical models~\cite{Krouppa:2015yoa,Emerick:2011xu}.

\section{Quarkonium in \mbox{p-A} or \mbox{d-A} collisions}
\label{sec:pA}
To correctly quantify the influence of hot matter effects, a precise knowledge of the cold matter effects is needed.
This study can be addressed thanks to the \mbox{d-Au} and the very recent \mbox{p-A} runs at RHIC energies (\sqrtsNN=200 GeV) or the LHC \mbox{p-Pb} run at \sqrtsNN= 5.02 TeV. Similarly to what is done to quantify how the medium affects quarkonium production in \mbox{Pb-Pb} collisions, cold nuclear matter effects are usually investigated through the nuclear modification factor \rpa ($R_{\rm dAu}$). 

At LHC energies, \jpsi\ \rpa has been studied by both ALICE~\cite{Abelev:2013yxa,Adam:2015iga,Adam:2015jsa,Leoncino_QM15} and LHCb~\cite{Aaij:2013zxa,Yang_QM15} at forward and backward $y$, down to \pt=0, and by ATLAS~\cite{Aad:2015ddl,ATLAS:Conf23,Hu_QM15} in the central rapidity region, focussing, in this case, only on the high-\pt\ \jpsi\ production.
The \jpsi\ \rpa shows a strong dependence on rapidity and \pt. While at backward-$y$ (correponding to the Pb nucleus flight direction), \rpa is close to unity, with a rather flat trend versus \pt, at forward-$y$ (corresponding to the proton flight direction) an increase towards high-\pt\ is observed. At mid-$y$ the \rpa shows a small \pt\ dependence, with a clear flattening in the high \pt\ range covered by ATLAS. 
\begin{figure}[h]
\centering
\includegraphics[width=15cm,clip]{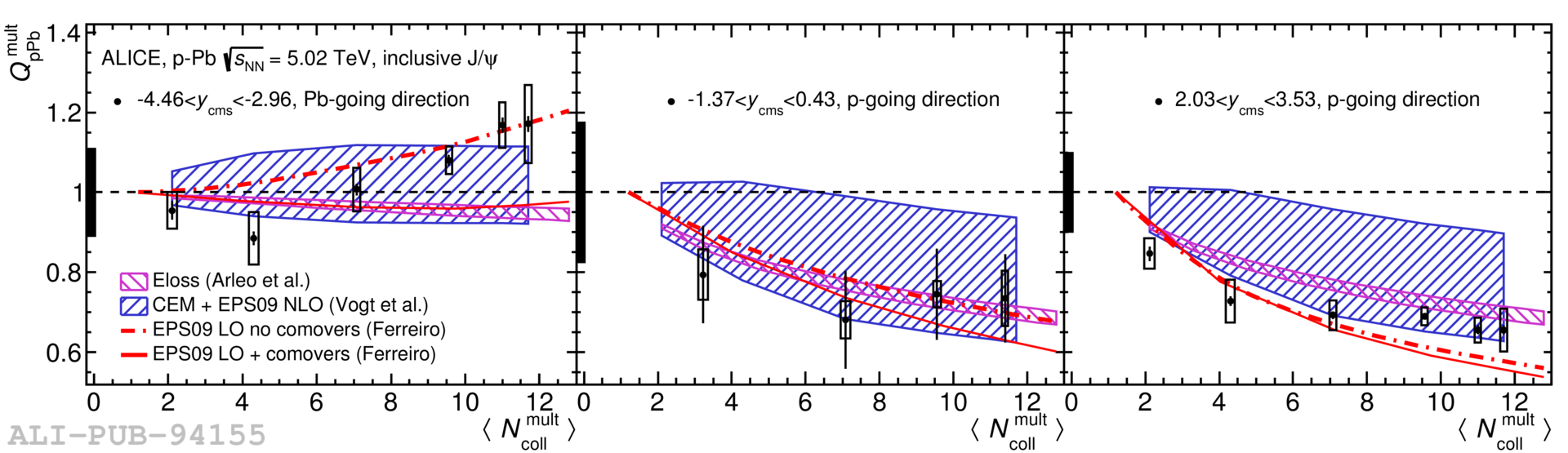}
\caption{\jpsi\ nuclear modification factor ($Q_{\rm pA}$), measured by ALICE~\cite{Adam:2015jsa}, as a function of centrality, at backward (left), mid (center) and forward rapidity. Results are compared with the theory models described in the text.}
\label{fig:centrjpsipA}       
\end{figure}

When studied as a function of centrality, the nuclear modification factor ($Q_{\rm pA}$) increases from peripheral to central collisions at backward rapidity, while at mid and forward-$y$, the \jpsi\ is more suppressed, with respect to \mbox{pp} collisions, in central events, as shown in Fig.~\ref{fig:centrjpsipA}. 

Results are within agreement with theory predictions, based on a pure nuclear shadowing scenario~\cite{Vogt:2010aa,Albacete:2013ei}, as well as partonic energy loss, either in addition to shadowing or as the only nuclear effect~\cite{Arleo:2013zua}. 

\begin{figure}[h]
\centering
\includegraphics[width=6.5cm,clip]{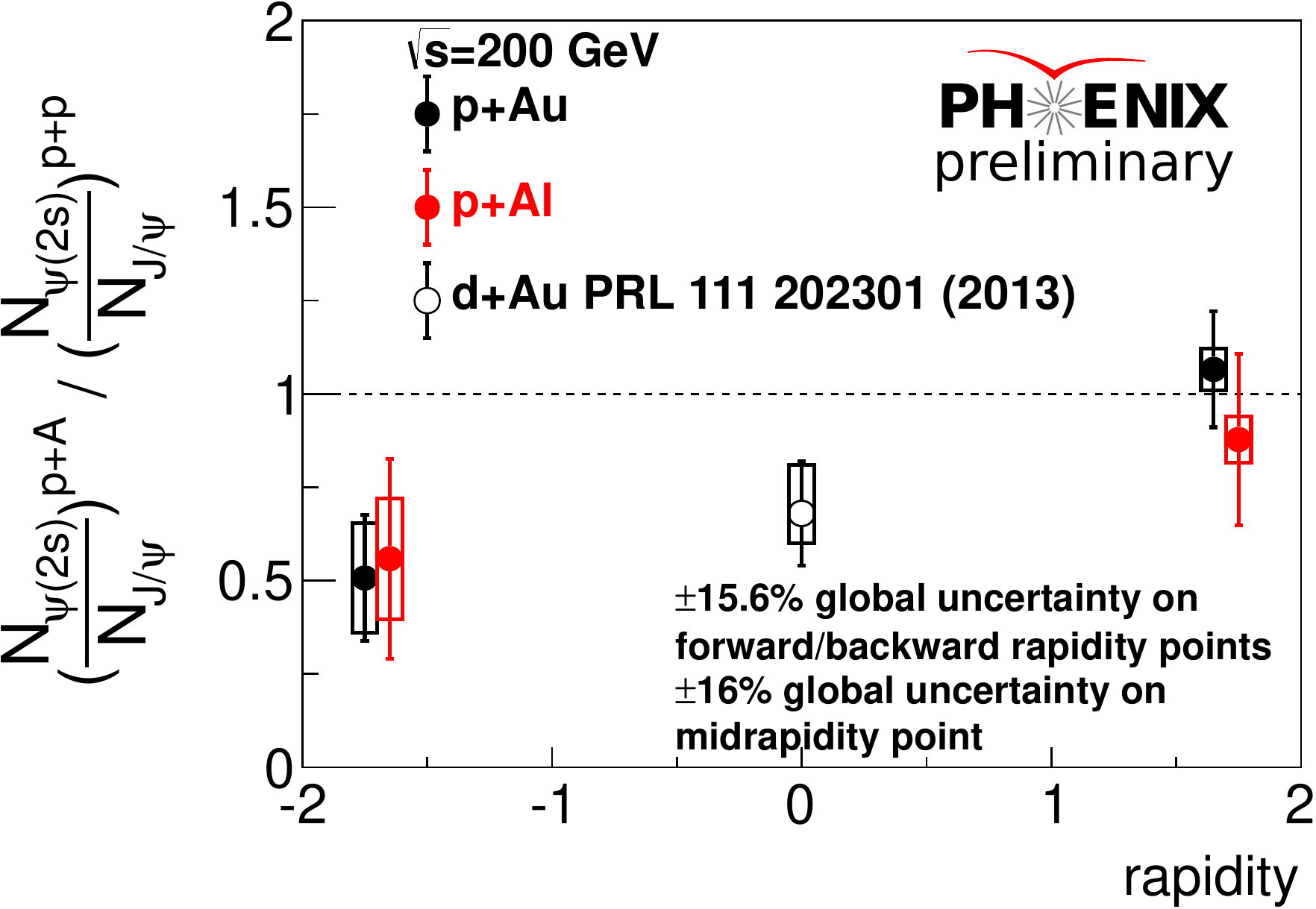}
\includegraphics[width=6.5cm,clip]{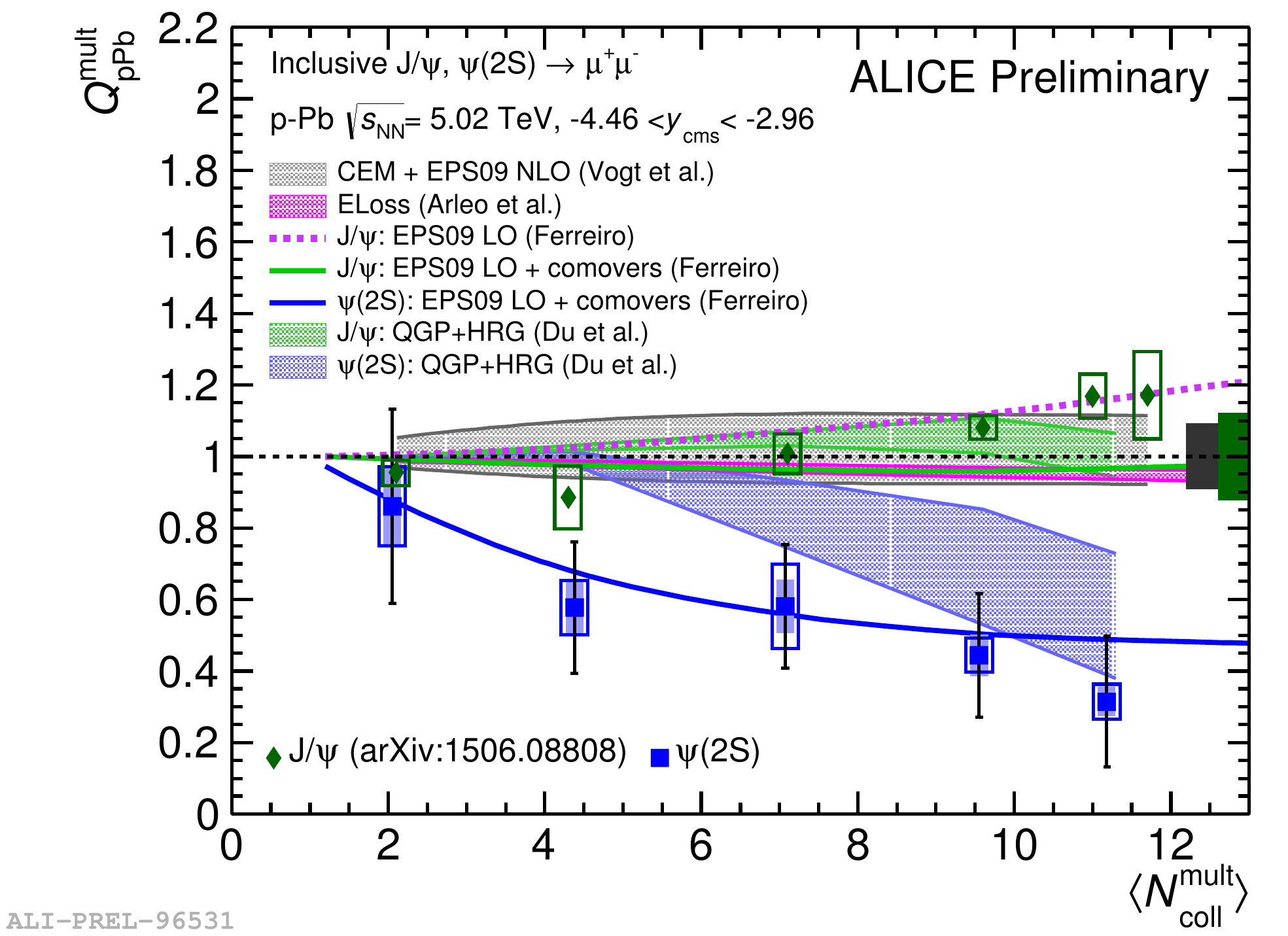}
\caption{Left: double \psip\ and \jpsi\ ratio in \mbox{p-A} (\mbox{d-A}) and \mbox{pp} as measured by PHENIX~\cite{Frawley_QM15}. Right: \psip \rpa measured by ALICE~\cite{Leoncino_QM15} at backward $y$, compared to the corresponding \jpsi\ one and to the theoretical calculations discussed in the text.}
\label{fig:psip_pA}       
\end{figure}

Being more weakly bound than the \jpsi, the \psip\ is a valuable probe to get further insight on charmonium behaviour also in \mbox{p-A} collisions.
At RHIC or LHC energies, the time spent by the $c\bar{c}$ pair in the created medium is much shorter that the time needed to the pair to evolve into a fully formed resonance state as the \jpsi\ or the \psip. Therefore, cold nuclear matter effects affect only the pre-resonant state and are expected to be very similar for the two charmonium states. 
However, at LHC~\cite{Abelev:2014zpa,Leoncino_QM15,LHCb:Conf5,Yang_QM15} and at RHIC, both in mid-$y$ \mbox{d-Au} and in backward-$y$ \mbox{p-A} collisions~\cite{Adare:2013ezl,Frawley_QM15}, the \psip\ suppression is unexpectedly found to be stronger than the \jpsi\ one, as shown in Fig.~\ref{fig:psip_pA} (left). 
The \psip\ nuclear modification factor, as a function of centrality~\cite{Leoncino_QM15}, shows an increase of the  suppression in most central collisions, with, in particular at backward-$y$, a rather different behaviour with respect to the the \jpsi\ one (Fig.~\ref{fig:psip_pA}). While the \jpsi\ behaviour can be explained in terms of nuclear shadowing or energy loss, an additional mechanism, affecting only the weakly bound \psip, is required to explain the observed pattern. A plausible explanation for the stronger suppression requires, therefore, the introduction of the \psip\ dissociation by comovers in a hadronic medium, created in \mbox{p-A} collisions with or without the existence of a short-lived QGP phase~\cite{Ferreiro:2014bia,Du:2015wha}. 

\begin{figure}[h]
\centering
\includegraphics[width=7cm,clip]{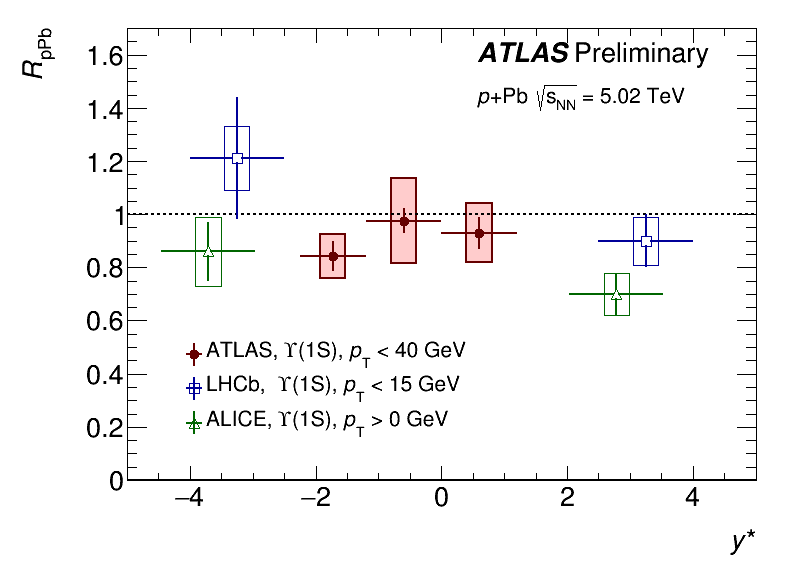}
\caption{\upsi(1S) nuclear modification factor as a function of $y$, as measured by ATLAS~\cite{ATLAS:Conf50,Hu_QM15}, ALICE~\cite{Abelev:2014oea} and LHCb~\cite{Aaij:2014mza}.}
\label{fig:upsi_pA}       
\end{figure}

Finally, \upsi\ results in \mbox{p-Pb} collisions have also been obtained by ATLAS~\cite{ATLAS:Conf50,Hu_QM15}, ALICE~\cite{Abelev:2014oea}, LHCb~\cite{Aaij:2014mza} and CMS~\cite{Chatrchyan:2013nza}. In the rather large rapidity range covered by these experiments, no clear $y$ dependence for the \upsi(1S) state is observed, as shown in Fig.\ref{fig:upsi_pA} (left). The centrality dependence is also rather flat and the \rpa is consistent with unity, indicating weak cold nuclear matter effects affecting the \upsi(1S). 
The excited \upsi\ states, measured by CMS~\cite{Jo_QM15} show a stronger suppression with respect to the \upsi(1S), suggesting final states at play, as also discussed in the charmonium sector. The \pt\ and $y$ dependence of this effect is feeble, as measured by ATLAS~\cite{Hu_QM15}. 

\section{Conclusions}
\label{sec:concl}
The investigation of the quarkonium production in heavy-ion collisions, already addressed by the SPS and RHIC experiments, is now enriched by a large wealth of results from the LHC experiments. These results, which refer to  
several charmonium and bottomonim states in different, but complementary, kinematic regions allow a deeper understanding of the quarkonium behaviour in \mbox{A-A} interactions. 
The new results presented at the Quark Matter 2015 Conference confirm the existence of two competing processes 
affecting the quarkonium production in nucleus-nucleus collisions: the suppression in the deconfined medium and the (re)combination of ${\rm Q}$ and $\overline{\rm Q}$ quarks. 
A deeper understanding of hot medium effects requires the knowledge of the underlying effects related to cold matter and present also in \mbox{p-A} (\mbox{d-Au}) interactions. The investigation of the \jpsi\ behaviour in proton-nucleus collisions has shown that several effects are at play, as nuclear shadowing and energy loss, while comover-like additional mechanisms seem to affect the excited quarkonium states. 
The LHC Run-II, with a higher center of mass energy and a larger luminosity, is going to provide even more precise results, addressing also measurements which were challenging in Run-I. These new data will allow us to sharpen our knowledge, aiming to a coherent description of quarkonium production in a hot and deconfined medium, from low to very high energies.





\bibliographystyle{elsarticle-num}
\bibliography{arnaldi_bib_QM15}







\end{document}